\documentclass{aa}
\input psfig.sty
\input epsf.sty

\def \msun {\hbox{$M_{\sun}$}}
\def \lsun {\hbox{$L_{\sun}$}}
\def \xmm {{\it XMM-Newton} }

\def \h5-1 { $h_{50}^{-1}$}

\begin{document}
%


   \title{The dynamical state of the Coma cluster with XMM-Newton }

   \titlerunning{The dynamical state of the Coma cluster}

   \author{D.M. Neumann \inst{1} 
\and
D.H. Lumb \inst{2}
\and
G.W. Pratt \inst{1}
\and
U.G. Briel \inst{3}
}

   \offprints{D.M. Neumann, ddon@cea.fr}

\institute{CEA/DSM/DAPNIA Saclay, Service d'Astrophysique,
L'Orme des Merisiers B\^at 709., 91191 Gif-sur-Yvette, France
\and
Science Payloads Technology Division, Research and Science Support
Dept., ESTEC, Postbus 299 Keplerlaan 1, 2200AG Noordwijk, Netherlands 
\and
Max-Planck Institut f\"ur extraterrestrische Physik, 
Giessenbachstr.,
85740 Garching,
Germany}

  \date{Received xxxxx; accepted xxxxx}

\abstract{
We present in this paper a substructure and spectroimaging study of the Coma 
cluster of 
galaxies based on \xmm data. \xmm performed a mosaic of observations of Coma 
to ensure a large coverage of the cluster. We add the different 
pointings together and fit elliptical beta-models to the data. We subtract
the cluster models from the data and look for residuals, which can be
interpreted as substructure. We find several
significant structures: the well-known subgroup connected to NGC~4839 in the 
South-West of the cluster, and another substructure located between NGC~4839 
and the centre of the Coma cluster. Constructing a hardness ratio image, which 
can be used as a temperature map we see that in front of this new structure 
the temperature is significantly increased (higher or equal 10~keV). We 
interpret this temperature
enhancement as the result of heating as this structure falls onto 
the Coma cluster. We furthermore reconfirm the filament-like structure 
South-East of the cluster centre. This region is significantly cooler than the 
mean cluster temperature. We estimate the temperature of this structure to
be equal or below 1~keV.
A possible scenario to explain the observed features is  
stripping caused by the infall of a small group of galaxies located around
the two galaxies NGC~4921 and NGC~4911 into the Coma cluster with a non-zero
impact parameter. We also see significant X-ray depressions North and 
South-East of NGC4921, which might either be linked to tidal forces due to
the merger with the Western structure or connected to an older cluster
merger.

}

\maketitle


\section{Introduction}

The Coma cluster, also known as Abell 1656, is one of the best studied clusters
in our Universe (see Biviano \cite{biviano} for a historical review). It has 
been 
explored in all wavelengths from radio to hard X-rays. However, despite the 
large number of observations, the 
physics of the Coma cluster and its dynamical state is not yet 
completely understood. The cluster hosts a powerful radio halo (Feretti \& Giovannini \cite{feretti}) and Beppo-Sax data recently revealed the existence of  
hard non-thermal X-ray emission (Fusco-Femiano et al. \cite{fusco-femiano}).

Extensive work has been carried out to measure the velocity dispersion and
distribution of the galaxies located in the Coma cluster. Colless \& Dunn 
(\cite{colless}) for example published a catalog of 552 redshifts and determined the
mean redshift of the cluster to be $cz = 6853$~km/sec and a velocity dispersion
of $\sigma_{cz} = 1082$~km/sec. Colless \& Dunn (\cite{colless}) find and confirm
the presence of substructure in the galaxy distribution: there is a sub-group
located around NGC~4839, which lies to the South-West of the cluster. The
existence of this substructure was originally found by Mellier et al. (\cite{mellier}) 
and Merritt \& Trimbley (\cite{merritt}). Furthermore, it was noted by 
Fitchett \&  Webster (\cite{fitchett}) that the two 
central galaxies, NGC~4889 and NGC~4874 have a significant difference in 
velocity, which is indicative of a recent merger. It is probable that 
NGC~4889 belonged
to a subgroup, which recently merged with the main
cluster and that NGC~4874 is the
central dominant galaxy of the main cluster.

There are a number of papers addressing the thermal X-ray emission of
the Coma cluster which comes from the hot thermal intracluster medium (ICM). White, 
Briel \& Henry (\cite{white}) discussed substructures in the Coma cluster based on 
ROSAT-data. Briel, Henry \& B\"ohringer (\cite{briel92}) analyzed the ROSAT 
All-Sky-Survey data and determined the X-ray surface brightness profile of the 
cluster out to a radius of roughly 100 arcmin. Hughes (\cite{hughes}) performed
an extensive mass analysis on the Coma cluster based on X-ray and optical
data.
Later work by Vikhlinin,
Forman \& Jones (\cite{vikhlinin}) revealed the existence of a filament-like substructure,
which is most likely linked to NGC~4911, a bright spiral galaxy located 
South-East with respect to the centre of the cluster. Spectroimaging
studies based on ASCA-data constrained the X-ray temperature 
distribution of the ICM. Honda et al. (\cite{vikhlinin}) and Watanabe et al. 
(\cite{watanabe}) 
presented temperature maps of the Coma cluster based on ASCA-data, which 
clearly showed that the ICM is not isothermal. Furthermore, Donnelly et al. 
(\cite{donnelly}) saw indications for a hot spot North of the centre of the Coma 
cluster. 
Recently, first results of the Coma cluster based on XMM-Newton data were 
published: Briel et al. (\cite{briel01}) discussed the overall morphology of the cluster and showed the existence of several point-like sources, which are 
linked to individual galaxies in the cluster and which have not been 
accounted for in previous studies. Arnaud et al. (\cite{arnaud}) presented a 
temperature map of the central part of the cluster based on \xmm data
with unprecedented spatial resolution. This map showed again variations of the
X-ray temperature (see Jansen et al. \cite{jansen} for an overview of \xmm
and Turner et al. \cite{turner} and Strueder et al. \cite{strueder} for an overview of the EPIC detectors). However, the
previously found hot spot found by Donnelly et al. \cite{donnelly} was not
confirmed.
Lastly, Neumann et 
al. (\cite{neumann}) presented strong evidence based on \xmm data that the 
substructure located around NGC~4839 (see above) is on its first infall onto 
the cluster from the South-West and has not yet passed 
its centre.  This is in contradiction with previous work by Burns et al. 
\cite{burns}, who suggested that the group around NGC~4839 has already passed
the centre of the Coma cluster once.

The aim of this paper is to perform a substructure and spectroimaging 
study of the Coma 
cluster based on \xmm data, to better understand the 
underlying physics. The work presented here is our first attempt to discuss the
overall dynamics of the cluster, and is by no means exhaustive. It aims to
show the large potential provided by the data obtained with \xmm .
The paper is organized as follows: after the 
introduction we present the observations in Sec.2. This is followed by the 
description of our data treatment in Sec.3. In Sec.4 we present our spatial 
analysis and in Sec.5 we describe our spectro-imaging results. In Sec.6 we
discuss our results and finish in Sec.7 with the Conclusion.
 
Throughout this paper, we assume $H_0=50$~km s$^{-1}$ Mpc$^{-1}$, 
$\Omega=1$ and
$q_0 = 1/2$. At a redshift of $z=0.0232$, $1^\prime$ corresponds to 38.9~kpc. 
 The cluster is sufficiently close so that variations of the geometry of the universe do not matter.


\section{Observations}

The size of the Coma cluster (roughly 3 degrees in diameter) greatly exceeds
the field-of-view (FOV) of \xmm. Therefore the observations of Coma
were performed in mosaic mode with several exposures pointing at different
regions of the cluster. The 
summary of all observations is shown in Tab.\ref{tab:expo}. The exposure times 
indicated are the effective observing times after flare rejection. We apply 
the same method and thresholds for flare rejection as described in 
Majerowicz, Neumann \& Reiprich (\cite{majerowicz}) for the pn and MOS-cameras.
During the observations the pn-camera was mostly
operating in extended-full-frame
mode.
There was one additional Coma pointing, in which only the pn-camera was active 
(Coma 2), for which we were not able to run the pipeline processing at the 
time of the analysis. Thus we discard this observation in the following.

\begin{table}
\begin{tabular}{cccccc}
\hline
pointing & camera & exposure & RA  & Dec. & date \\
no. & & in sec & in deg. & in deg. & mm/yy \\
\hline
\hline
10 & m1 &  18576 & 194.910 &  28.1278 & 06/00 \\
   & m2 & 18584 & & & \\
   & pn & 17500 & & & \\
11 & m1 & 13845 & 194.652 &  28.3989 & 06/00\\
   & m2 & 15237 & & &\\
   & pn & 13754 & & &\\
1  & m1 & 26559 & 194.199 & 27.4019 & 06/00\\
   & m2 & 26594 & & &\\
   & pn & 23732 & & &\\
2  & m1 & 26046 & 194.427 &  27.7272 & 12/00\\
   & m2 & 26062 & & &\\
   & pn & 15000 & & &\\
3  & m1 & 12519 & 194.634 &  27.4033 & 06/00\\
   & m2 & 11523 & & &\\
   & pn & 11730 & & &\\
4  & m1 & 476   & 195.019 &  27.5233 & 06/00\\
   & m2 & 582   & & &\\
   & pn & 2970  & & &\\
5  & m1 & 14977 & 194.865 & 27.7814 & 05/00\\
   & m2 & 16485 & & &\\
   & pn & 14211 & & &\\
6II& m1 & 14943 & 194.708 & 27.9811 & 12/00\\
   & m2 & 14942 & & &\\
   & pn & 8874  & & &\\
6  & m1 & 6548  & 194.708 & 27.9811 & 06/00\\
   & m2 & 6744  & & &\\
   & pn & 3969  & & &\\
7II& m1 & 23755 & 194.365 &  28.1447 & 12/00\\
   & m2 & 23245 & & &\\
   & pn & 8874  & & &\\
7  & m1 & 3153  & 194.365 & 28.1447 & 06/00\\
   & m2 & 2960  & & &\\
   & pn & 358   & & &\\
8  & m1 & 23455 & 195.357 & 27.7314 & 12/00\\
   & m2 & 24172 & & &\\
   & pn & 18430 & & &\\
9  & m1 & 19951 & 195.136 & 27.9497 & 06/00\\
   & m2 & 20361 & & &\\
   & pn & 15443 & & &\\
bgd& m1 & 21145 & 195.404 & 27.3311 & 06/00\\
   & m2 & 21563 & & &\\
   & pn & 17534 & & &\\
cen& m1 & 16103 & 194.945 &  27.9500 & 05/00\\
   & m2 & 15894 & & &\\
   & pn & 14595 & & &\\
\hline
\end{tabular}
\caption{The summary of exposure times and coordinates for the different 
Coma pointings.}
\label{tab:expo}
\end{table}

\section{Data treatment}

Our analysis presented here is based on SAS-version 5.1 and concerns only
events in the field-of-view (FOV).

\subsection{Vignetting correction}

In order to correct for vignetting we apply the photon weighting method 
described in Arnaud et al. (\cite{vignetting}). To outline it briefly: to each observed 
event at energy $E$ falling onto the detector position $x,y$
we attribute a correcting factor: 
\begin{equation}
\omega = \frac{\mbox{effective area}(x=0,y=0,E)}{\mbox{effective area}
{(x,y,E)}}
\end{equation}

the effective area takes into account the mirror surface, the RGA structure
(in the case of the MOS-cameras), the filter transmission as well as the 
quantum efficiency of the detectors. We assume in our anaylsis that the filter 
transmission and the quantum efficiency are flat over the entire field of the
detector. $x=0, y=0$ stand for on-axis position. 

All images presented in this study are vignetting corrected. For this we 
create images by adding the different weight factors of the photons into 
image pixels. The statistical uncertainty per image
pixel is calculated as 
\begin{equation}
\sigma =  \sqrt{\Sigma_{i=1}^{n} \omega^2} 
\end{equation}
where $n$ is the number of photons per pixel.

\subsection{Merging of the different exposures}
\label{sec:merging}

While merging the data from different exposures we observed certain offsets 
of the
order of a few arcseconds for which we corrected by looking at point sources, 
which were visible in two or more pointings. We shifted the pointings 
so that the same point source, visible in different exposures, appeared in 
each pointing at the same sky coordinate position. We estimate the remaining
errors in the order of a few arcseconds.

\begin{figure*}
\psfig{figure=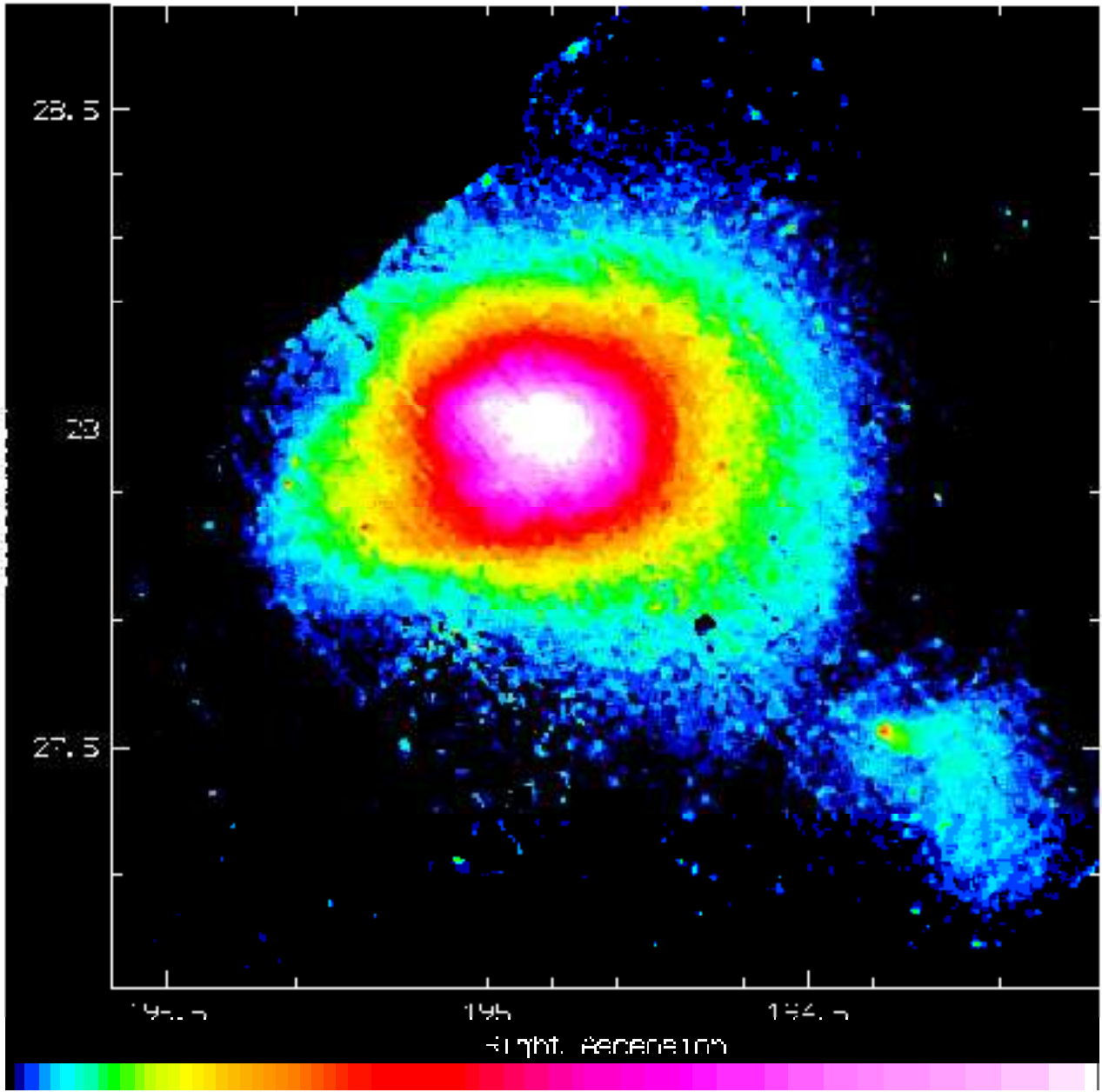,width=17cm}
\caption{ The Coma cluster in the 0.5-2.0~keV band. The image is the sum of
all the different exposures (see Tab.\ref{tab:expo}) of the different MOS
and pn cameras (see text). The image is background and vignetting corrected.
We applied a small median filter to mask out effects caused by 
detector gaps.  Colour scale: dark blue regions correspond to countrates of 0.016 cts/sec/arcmin$^2$ and white regions to countrates $\geq$ 0.288cts/sec/arcmin$^2$.}
\label{fig:merge}
\end{figure*}

Fig.\ref{fig:merge} shows the resulting countrate image in the 
energy band 0.5-2.0~keV after merging the different observations 
together (MOS and pn) and subtracting the background. For the background we
 used the Blank-Sky observations provided by D. Lumb and retrievable from the 
XMM-Newton web-pages. The background subtraction is not important for the
construction of the countrate images, however, it is important for the
determination of the hardness ratio map (see below). 

In order to obtain Fig.\ref{fig:merge} we perform the following data 
treatment:

\begin{itemize}

\item Merge the photon weight images of the individual MOS (MOS1+MOS2) 
pointings together (image: {\it mergemos}) in the energy range 0.5-2.0~keV

\item Create the corresponding MOS mosaic exposure map (image: {\it 
mergemosexpo}), which consists of the 
sum of the different detector masks\footnote{The 
detector mask represents the geometry of the detector which includes 
gaps between the
CCD's, the region of FOV, and excludes bad pixels, which were identified in 
the pipeline processing} per observation and camera (MOS1 and MOS2), which 
are multiplied by the observing time of the corresponding exposure before
summing.

\item Divide {\it mergemos} by {\it mergemosexpo} = {\it countratemos}

\item Create corresponding background countrate image for MOS (image:
{\it backgroundmos})
in the same 
energy range as in {\it mergemos}. For each MOS 
exposure we construct a vignetting corrected background exposure, for 
which we use as input the  blank sky observations compiled by D. Lumb. The 
corresponding photon weight images are added together. We construct the 
corresponding exposure map of the background and divide the photon weight image
by the exposure map. The result is a background countrate image 
{\it backgroundmos}.

\item {\it countratemos} - {\it backgroundmos} = {\it sourcemos}

\item Merge the photon weight images of the pn images (image: {\it mergepn}) 
in the same energy range as was used for MOS.

\item Create the corresponding pn exposure map (image: {\it mergepnexpo}) in 
the same way as done for MOS

\item Create background countrate image for pn in the same manner as for MOS
(image: {\it backgroundpn})

\item Divide {\it mergepn} by {\it mergepnexpo} = {\it countratepn}

\item {\it countratepn}  - {\it backgroundpn} = {\it sourcepn}

\item {\it sourcemos + sourcepn} = {\it total background-subtracted 
countrate image} = Fig.\ref{fig:merge}

\end{itemize}

For all observations MOS and pn were operated at the same time and had, 
perhaps
 with the exception of pointing no.4, sufficiently long exposure times 
 to give for each type of instrument an adequate countrate estimate per pixel. 
Therefore it is correct to add together the different 
countrate images obtained by the different detectors. If for one 
observation only one type of camera had been available
we could not add the countrate images since this individual exposure would
create biases in our mosaic image due to the different sensitivities and
spectral responses of the detector types. 

We want to stress that the Coma pointings were performed in 
medium filter mode while the data used for the background are compiled from 
observations mainly using thin filter. However, since we only use
photons at energies above 0.5 keV we expect the difference of the different
filters to be negligible. Furthermore, at regions with high signal-to-noise
ratio we expect that the background subtraction is not important.

\section{Spatial analysis}

\subsection{Substructure analysis}

Our aim in this study is to look for substructure and temperature variations
in order to better understand the dynamical state of the Coma cluster.

For measuring substructure we  fit an elliptical beta-model to the 
cluster images, which describes a relaxed cluster without substructure 
. Subsequently, we create a synthetic
cluster image based on the best fit parameters of the beta-model. 
This model is subtracted from the data. The residuals remaining after 
subtraction, which indicate substructure, are examined for their significance.
This method has been applied to useful effect in several other studies, e.g., 
 Neumann \& B\"ohringer \cite{neumann97}, \cite{neumann99} and 
Neumann \cite{neumann99b}.

In the following we describe the different steps of our substructure analysis
in detail.

\subsubsection{Beta-model}

In fitting elliptical beta-models to the XMM-Newton data, we take as 
input the vignetting corrected images in the 0.5-2.0 keV energy band. This
energy range was chosen since it provides the best signal-to-noise 
ratios (see Scharf \cite{scharf}). For this analysis we did not 
subtract the background, since we include it as fit parameter. We compare 
(see below and Tab.\ref{tab:back})
the fitted background values with the countrate values observed in the 
background observations. This is a consistency check for the 
applied background subtraction.

We encounter small number Poisson statistics in the pixels located at the 
outskirts of the cluster. This produces non-symmetric error bars for
the pixel values. Since we use $\chi^2$ statistics for our beta-model fitting, 
which assumes a symmetric Gaussian probability distribution for the photons, 
we have to correct for the asymmetry of the error bars. An easy 
way to do this is to apply a Gaussian filter ($\sigma=15^{\prime\prime}$) 
to the image before fitting the model. This procedure has proven to be valid 
and gives good results as shown in previous studies (e.g. Neumann \&
B\"ohringer \cite{neumann97}, \cite{neumann99}; Neumann \cite{neumann99b}). 
In order to account properly
for the vignetting, in our calculation of the error bars per pixel  
we quadratically add the different weighting factors for each event in each
 pixel before Gaussian filtering (see eq(2)), and use this image as estimate 
for the uncertainties in each pixel.

To check for possible instrument-dependent variations we treat the MOS- and 
the pn-data separately. The fit results are displayed in Tab.\ref{tab:beta}.
They are in good agreement with each other and with previous studies, like 
Briel, Henry \& 
B\"ohringer (\cite{briel92}) . Since we are only interested in the best-fit model
in order to subtract it from the cluster data and check for residuals, we
do not determine error bars on the fit parameters, which would be 
time consuming since the error bars would have to be estimated via a 
Monte-Carlo approach.

The centre of both the MOS and the pn-data fit coincide relatively well with
the position of the central galaxy NGC~4874, at: 
RA=194.898 and Dec.=27.963. The distance between the model centre and this
galaxy is less than 2 arcmin.

\begin{table}
\begin{tabular}{ccc}
\hline
 & MOS & pn \\
\hline
\hline
observed background & $1.2\pm 0.3$ & $2.9\pm 1.1$  \\
countrate in $10^{-4}$counts/sec/pixel & & \\
\hline
fitted background & $1.2$ & $3.8$ \\
countrate in $10^{-4}$ counts/sec/pixel & & \\
\hline
\end{tabular}
\caption{The observed and fitted background values of the Coma exposures. 
The observed background is determined from the blank-sky-field observations. 
The fitted background value is the result from the elliptical beta-model 
fitting. The error bars are 68\% confidence level.
} 
\label{tab:back}
\end{table}

\begin{table*}
\begin{tabular}{ccccccc}
\hline
camera & $S_0$ & $x_0$ & $y_0$ & $r_{c maj}$ & $r_{c min}$ & $\beta$ \\
 & cts s$^{-1}$ arcmin$^{-2}$ & in deg. (J2000) & in deg. (J2000) & 
in arcmin & in arcmin & \\ 
\hline
\hline
MOS & 0.094 & 194.929 & 27.9586 & 11.1 & 9.0 & 0.72 \\
pn & 0.28 & 194.927 & 27.9597 & 12.1 & 11.2 & 0.85 \\
\hline
\end{tabular}
\caption{The parameters from the elliptical beta-model fit. $S_0$ is the
central intensity, $x_0$ and $y_0$ are the central coordinates in RA. and Dec.
respectively. $r_{c maj}$ and $r_{c min}$  are the two core radii in major and
minor axis direction. }
\label{tab:beta}
\end{table*}

\subsubsection{Significance of residual map after beta-model subtraction}

\begin{figure*}
\psfig{figure=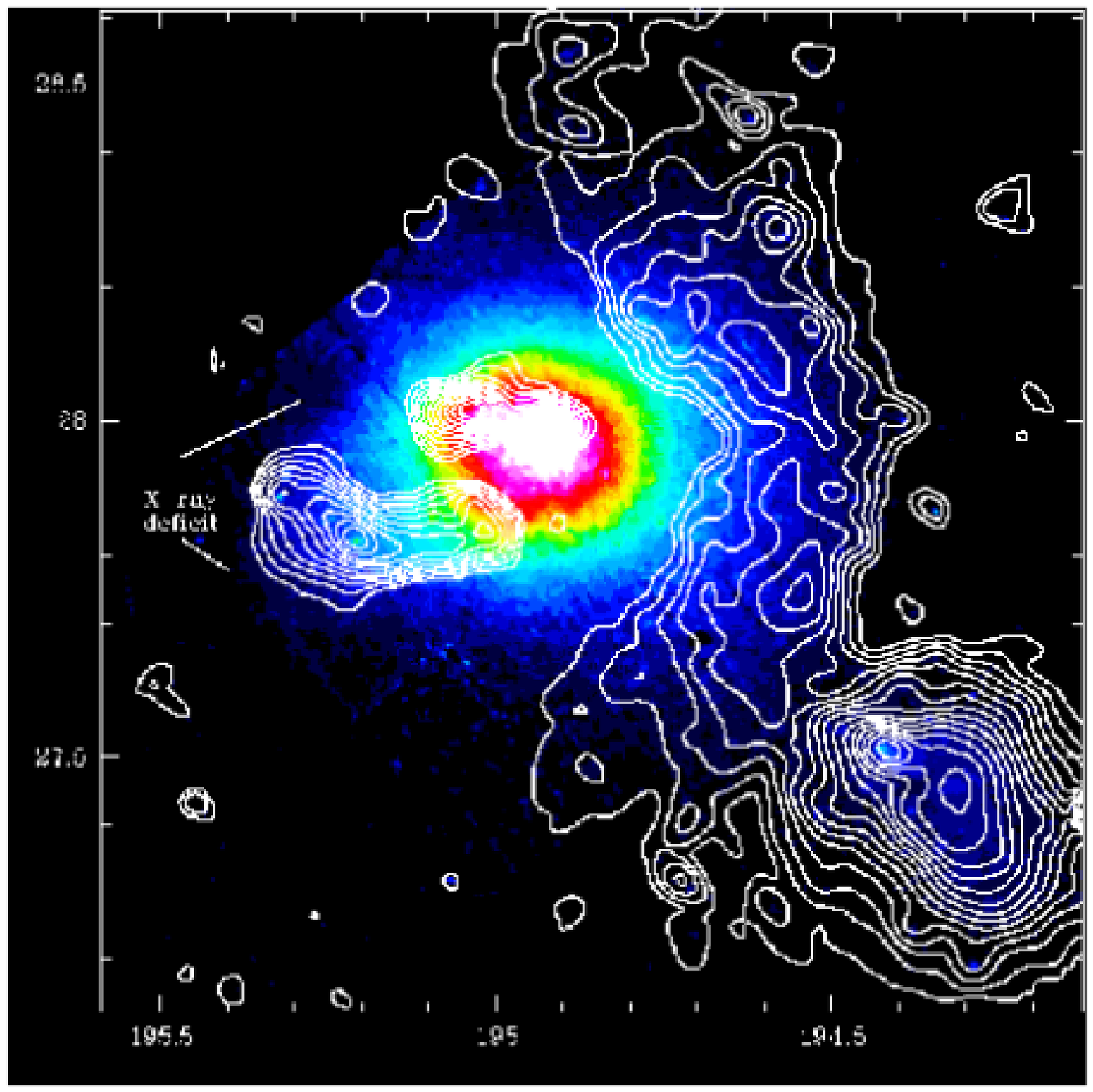,width=17cm}
\caption{The image is the same as in Fig.\ref{fig:merge} only with different
image value cuts and a linear look-up-table (lut) instead of a logarithmic lut.
The contours are the significances of the residuals above main relaxed
cluster and background emission (see also text). The lowest contour and the 
step width between two contours are each 5 $\sigma$.  Colour scale: darkblue regions correspond to a countrate of 0.0192 cts/sec/arcmin$^2$ and white regions to countrates $\geq 0.32$ cts/sec/arcmin$^2$.}
\label{fig:resi}
\end{figure*}

We subtract the cluster model based on the best fit parameters of the 
beta-model and calculate the corresponding significance of the 
residuals for each instrument type independently. The exact method we use for 
calculating
the significances is described in Neumann \& B\"ohringer 
(\cite{neumann97}). 
For our residual calculation we apply a Gaussian filter with $\sigma=1^\prime$.
We obtain two significance maps, one for the MOS-cameras, and the other 
for the pn-camera. The residuals obtained for each type of camera are very
similar.
We quadratically add the two different significance maps 
in order to obtain a residual image with better statistics. The result is 
shown in Fig.\ref{fig:resi}. The image over which we 
display the significance contours is the same as in Fig.\ref{fig:merge}.

We want to stress that the resulting residuals with a significance equal or 
higher than +10 $\sigma$ are extremely stable against changing the parameters 
of the subtraced model. 

\subsubsection{Results}

The residual map shown in Fig.\ref{fig:resi} shows several significant structures:

\begin{itemize}

\item the extended X-ray emission around and behind NGC~4839, which was 
already detected and described in previous papers. 

\item a significant (the lowest contour is $5\sigma$) structure in
the West, which is located between the Coma centre and the subgroup around 
NGC~4839 at a distance of about 0.9~Mpc. The structure is 
elongated in North-South direction with an extent of about 50 arcmin, which 
corresponds to about 1.7~Mpc. The width of the
structure in East-West direction is about 20 arcmin or 0.6~Mpc. The structure
shows two extended maxima at RA=12:58:1.26 and  Dec.=27:44:32.25 for
the first maximum and RA=12:58:16.88 and  Dec.=28:6:17.68 for the second. The 
entire structure is slightly bent tangentially to the Coma centre.

\item a residual with a filamentary shape is found to the East of the cluster centre.  A similar residual at the same location was already 
presented by Vikhlinin et al. (\cite{vikhlinin}) in a wavelet analysis of 
ROSAT data. The structure is elongated in the East-West direction 
and bends to the North at its Eastern end. 
The projected length of this filamenary structure is in 
the order of 28 arcminutes (which corresponds to roughly 1~Mpc) and has a 
width in the order of 6 arcminutes (about 200~kpc).

\item  a deficit in X-ray emission is detected North and South-East
of the filamentary Eastern structure (see above). The precise significance and
shape of these X-ray decrements is hard to determine (it is around 
-20$\sigma$), since it is very sensitive to the adopted background level, 
however, the signal is persistent. This deficit in 
X-ray emission can already be observed in Fig.\ref{fig:merge} comparing the 
overall emission South-East of the cluster with the overall emission North and 
West of the Coma cluster. We do not show contours of these negative residuals
in Fig.\ref{fig:merge} since their shape strongly depends on the adopted
background level.

\item an excess of
emission in the centre, which is linked to the two dominant central galaxies
NGC~4874 and NGC~4889. These residuals are signs for the two galaxies to have
deep individual gravitational potentials, which create small gas halos within
the larger scale of the cluster.

\end{itemize}

One important question is whether the detected residuals might not be 
an artefact due to improper modelling, for example the 
residuals could be due to the wrong determination of position angle 
of the subtracted model. In order to investigate 
this possibility we construct cluster models with different position
angles which are subsequently subtracted from the Coma data.
Even when we change the position angle by 90 degrees, the structures with 
positive residuals keep their 
overall morphology and significance. Since the centre 
of the model is very close to the central galaxy NGC~4874, it is most likely 
that the residuals cannot be caused by incorrect positioning of the centre of
 the subtracted model.

\subsection{Spectro-imaging analysis}

In order to measure the temperature distribution of the ICM
we determine the hardness ratio map for the Coma cluster. We make
use of Fig.\ref{fig:merge} and of the countrate image in the energy band 
2-5~keV.

Both of these images are constructed as described in Sect.~\ref{sec:merging} 
and are fully background-subtracted. Background subtraction is crucial to the 
construction of the hardness ratio map; without this correction 
background-dependent variations would appear in the final image.

 In order to 
avoid artefacts due to photon statistics, we apply
a median filter to the images in the 0.5-2.0~keV and the 2.0-5.0~keV band.
We filter over a region of 12 by 12 pixels (pixel size is $15^{\prime\prime} 
\times 15^{\prime\prime}$) before we divide the image obtained in the higher
energy band by the image in the low energy band. Applying a median filter not 
only helps for the statistics but also filters out the emission of 
individual sources such as galaxies or AGN. 
Since there are insufficient statistics in the outskirts of the cluster 
to reliably constrain the hardness ratio, 
we restrict ourselves to regions in which the countrate in the energy
band 2-5~keV band is above $2\times 10^{-4}$~cts/sec/pixel. 
This is a very 
conservative limit (see below) which helps avoid artefacts 
linked to imperfect background subtraction or to small number statistics.

\begin{figure*}
\psfig{figure=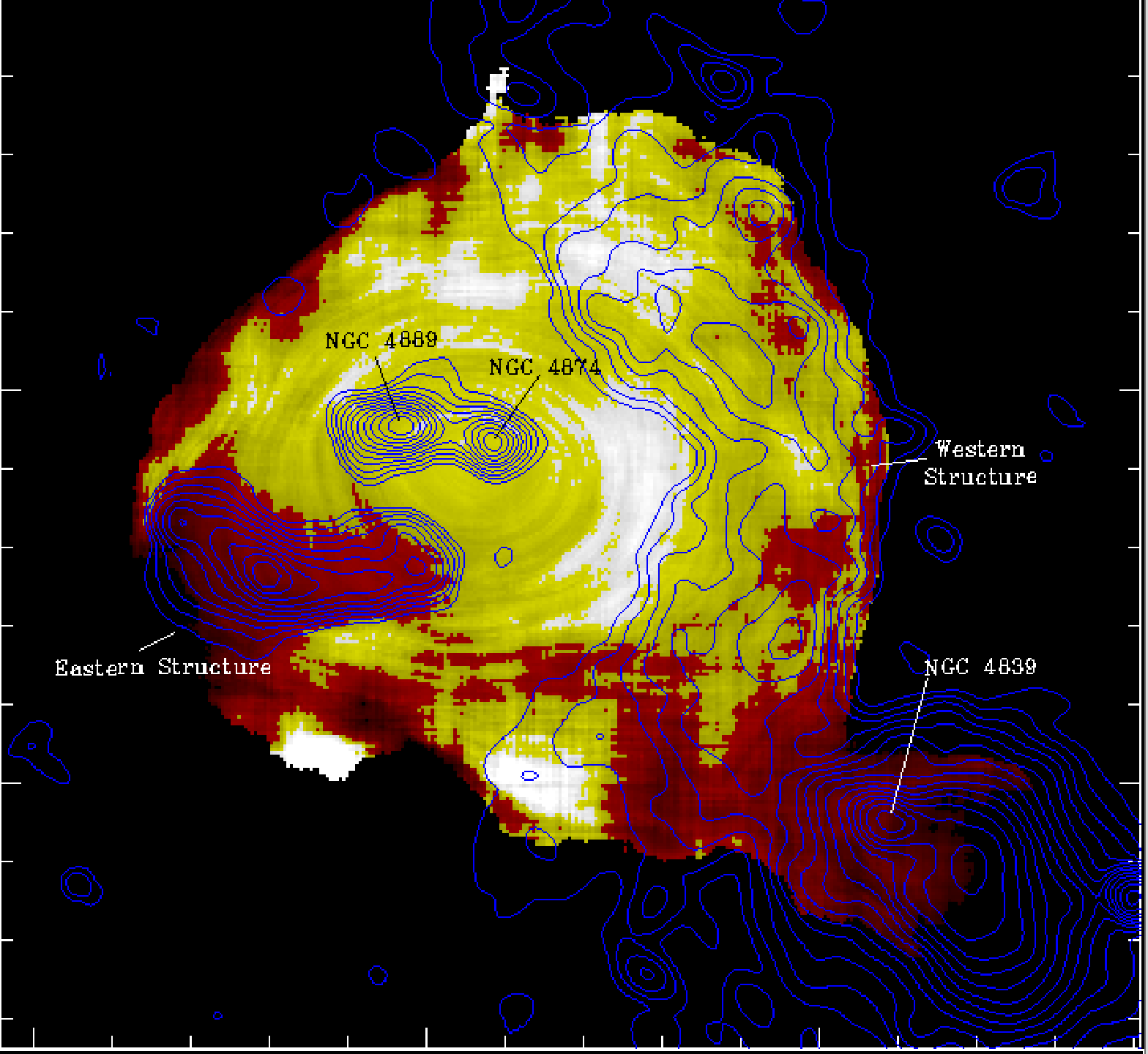,width=17cm}
\caption{The hardness ratio image obtained by dividing
the image in the 2-5~keV band by the image in the 0.5-2~keV band (see also
text). Assuming an isothermal spectrum, red regions correspond to parts of
the cluster with temperatures below 8~keV, yellow regions correspond to 
k$T>$8~keV and white regions have temperatures above 10~keV. The contours are 
the significance of the
residuals, see also Fig.\ref{fig:resi}.}
\label{fig:hr}
\end{figure*}

\subsubsection{Results}

The hardness ratio image is shown in in Figure~\ref{fig:hr}. The image is in three colours; red shows temperatures 
below 8~keV, yellow regions between 8-10~keV and white are regions with 
temperatures above 10~keV. Our results are in good agreement with those 
obtained by Arnaud et al. (\cite{arnaud}) and Briel et al. (\cite{briel01}).

Are the differences in hardness ratio
significant or not? In order to evaluate the errors, we perform the following
approximation: 

we only look at regions in the hardness ratio image in
which the countrate in the 2-5~keV band is larger or equal to
$2\times 10^{-4}$~cts/sec
per pixel (the pixel size is $15''\times 15''$). Assuming a mean exposure of
20~ksec per camera, we obtain 8 photons per pixel\footnote{this takes into
account the fact that we calculated the countrate assuming only one MOS camera
with a total exposure time of MOS1 plus MOS2.} . Since the observed mean
hardness ratio is roughly 0.3 (which corresponds to a plasma temperature of
about 8~keV), the countrate in the 0.5-2~keV band is a 
factor of about 3 higher than the countrate in the higher energy band. Thus, 
we obtain roughly 24 photons per pixel in a 20~ksec exposure. This calculation 
takes only into account the source countrate. For the proper error calculation 
we also have to include the background countrate which we subtracted via the 
blank sky observations. The total background countrate in the 0.5-2.0~keV band
is $4.1\times 10^{-4}$cts/sec/pixel and for the 2.0-5.0~keV band
$2.3\times 10^{-4}$cts/sec/pixel.
 Furthermore, for the
hardness ratio image we applied a median filter derived from the
pixel values in a window of $25\times25$ image pixels.
If we assume that the median filter has the same error as if one averages over
all pixels, a good approximation, we can determine the
errors by calculating the total number of photons and using error propagation
including the background. We obtain  a $1-\sigma$-error of 0.01 in the
hardness ratio. This corresponds to about 0.5~keV for temperatures below
10~keV. We want to stress that this
is the error estimate for the outer region. In the inner parts, where the
statistics are much better, the errors are much smaller. As an 
example: from the outer parts from the cluster, to the centre, the increase of 
X-ray intensity is of the order of a factor of 20. 

One final point concerning the results: the hardness ratio image shows 
two ``hot spots'' to the 
South. The significance of these two points is very low since they lie both
in pointing no.4, which has an exceptionally low exposure time (see 
Tab.\ref{tab:expo}). It is thus quite likely that these features are 
statistical artefacts.

\section{Discussion}

\subsection{Infall of a substructure located between Coma center and NGC~4839}

In Figure~\ref{fig:hr}, there is an obvious, large residual to the West of
 the cluster centre. The region lying between this structure and the 
Coma centre shows a temperature increase to temperatures 
equal to or above 10~keV. This suggests that 
it is heated via compression or via shock waves, which
were created during the infall of this sub-structure onto the Coma centre.
The creation of shock waves is observed in hydrodynamic simulations of galaxy 
cluster mergers (see for example Schindler \& M\"uller \cite{schindler}).  
It seems unlikely that the structure is a chance alignment in the 
line-of-sight. 

The two maxima in this western structure could 
be caused by a disruption of a galaxy group during its infall onto the cluster 
centre into several smaller units. It is relatively unlikely that two 
galaxy groups are falling onto the Coma cluster at exactly the same time.
 
Previous work has shown that the galaxy group around NGC~4839 is falling onto 
the Coma cluster (see for example Neumann et al. 2001 and 
references therein). The infall direction 
coincides with a filament which connects the Coma cluster with Abell 1367, and which is part of the `Great Wall'.  One may ask if the structure between NGC~4839 and the cluster centre is also falling in from the same direction. If a subcluster infalls from this direction, it will heat up the gas between it and the main cluster via compression. This will lead to an increase in the ram pressure encountered by the subcluster. Pressure increases linearly with temperature for an ideal gas  ($P=nkT$): the relative difference in temperature between the hot region and the main cluster is $\sim 20\%$, indicating a similar increase in pressure, and which may be sufficient to slow down the infalling gas. It is possible that, due to inertia, the gas in the subcluster is not abruptly slowed down, but instead may move in a direction where there is no increase in pressure. This movement could be tangential from South to North. In this case, the hot region to the North can naturally be explained as due to compression of the remaining gas in the subcluster.

The increase of temperature towards the North has been seen in ASCA-data (see Watanabe et al. \cite{watanabe} and Honda et al.\cite{honda}). It was interpreted as being due to heating by a shock wave created 
from a previous merger from a galaxy group located around NGC~4889, which is now one of the two main galaxies in the centre of the Coma cluster. It is not possible at present to distinguish between the two explanations. 

Examination of DSS plates does not show any obvious overdensity of 
galaxies coinciding with the Western residual structure. In this case, one might ask whether
this structure is the remaining gas from the subcluster which was once associated with NGC~4889. Ram pressure is known to act much more efficiently on gas than on galaxies, which would provide a natural explanation for the differences in location of the gas and the galaxy. The effectiveness of ram pressure stripping can be seen in the case of the subgroup around NGC~4839, in which the gas is clearly lagging behind the galaxies on the infall trajectory (Neumann et al. \cite{neumann}). Taking the projected distance between NGC~4889 and the western substructure ($\sim 1$ Mpc) and a typical cluster merger or sound crossing timescale of $10^9$ years, we can calculate the mean difference in velocity between the galaxy and 
the gas. This value is close to the velocity dispersion of Coma (in the order of 1000~km/sec - Colless \& Dunn \cite{colless}), which is not surprising given that typical cluster sizes are of order of a few Mpc. The feasibility of this scenario can only be assessed with high resolution hydrodynamic simulations, which take into account the differences of ram pressure between the galaxies and the gas of the merging subgroup.

\subsection{The residual to the East}

This compact region clearly shows lower values in the hardness ratio 
image, indicating that this part of the cluster is cooler than the 
mean cluster temperature. 

The filamentary form of the structure suggests that it may be linked 
to some
sort of infall process in connection with ram pressure stripping. Since ram
pressure stripping works efficiently only in the centre of the cluster, where
the gas density is high, we suppose that this structure is relatively close
to the centre of the cluster and not a projection effect. NGC~4911, a
bright galaxy which lies in the region of the substructure, shows a relative
difference in velocity with respect to the Coma cluster of 1000~km/sec, which 
is close to the velocity dispersion of the cluster as a whole. 

\subsubsection{Gas mass estimate of substructure}

To determine the origin of the filamentary Eastern 
structure, and more specifically to know whether the the observed stripped gas 
comes from an intragroup medium or from an interstellar medium linked to 
galaxies, we attempt to estimate the mass of gas of this structure. A high 
value 
for the mass would suggest that the gas comes from an intragroup medium, a 
low value would indicate that the  origin of the gas is from gas located
inside the galaxies.

The structure has a projected distance from the Coma
centre of roughly 1~Mpc and enhances the observed flux with respect to the 
fitted beta-model by about $30\%$. Using a length of 1~Mpc, a diameter of 
200~kpc, and assuming cylindrical symmetry (thus a thickness of the structure 
in the line-of-sight of 200~kpc), the volume of the substructure 
$V_s \sim 0.04~{\rm Mpc}^3$.

In order to calculate the gas mass of this structure we have to determine the
contributions of the main cluster gas and the gas in the 
substructure. The contribution of the main cluster gas has to
be integrated along the line-of-sight. For this we make use of the previously
determined betamodel parameters by Briel et al. (\cite{briel92}) ($\beta=0.75, r_c=0.42$~Mpc$,n_{e0}=2.9\times 10^{-3}$cm$^{-3}$) . The determination of the gas mass of the substructure depends on its location with respect to the cluster. If located at a large distance to the cluster we can simply add the two X-ray contibutions of main cluster and substructure. In this case we find an electron density of the subgroup of $n_{es}= 1.1\times 10^{-3}$ cm$^{-3}$. Integrating this density over the previously calculated volume we find a total gas mass for the structure of 
M$_{gas,s}=8\times 10^{11}$ \msun. If we assume that the structure is very close to the centre and that in fact the 
physical distance towards the Coma centre is close to its projected distance 
we need to take into account that the gas in the substructure is at the 
same position than the gas of the main cluster, which enhances the X-ray emission in this region\footnote{  If the two regions are only a chance alignment in the line-of-sight the X-ray emission ($xem$) in this region follows: $xem \propto n_{es}^2 + n_{em}^2$ where $n_{em}$ is the electron density of the main cluster gas. In the case that the two regions are located at the same place $xem$ behaves as: $xem \propto n_{etot}^2 = (n_{es}+n_{em})^2 = n_{es}^2 +2n_{es} n_{em} + n_{em}^2 > n_{es}^2+n_{em}^2$. The factor $2n_{es} n_{em}$ boosts the X-ray emission if the two regions are physically connected.}. 
In this case the required gas density for the subgroup is lower 
than in the case of no physical connection between the two components. We 
obtain in this case an electron density of 
$n_{es}=7.5\times 10^{-4}$cm$^{-3}$, 
and a gas mass of M$_{gas,s}=5\times 10^{11}$ \msun. Combining the two cases
as the two possible extremes we find that the gas mass of the substructure
lies between $5\times 10^{11}$ \msun $<$ M$_{gas,s}<8\times 10^{11}$ \msun. 
This value for the gas mass is at the low end for a group of galaxies 
(Mulchaey \cite{mulchaey}) and 
suggests that a part of the gas might also come 
from the interstellar medium of galaxies themselves.
The obtained values are in very good agreement with the value obtained by Vikhlinen et al. 
\cite{vikhlinin} with $5\times 10^{11}$\msun.

\begin{figure*}
\psfig{figure=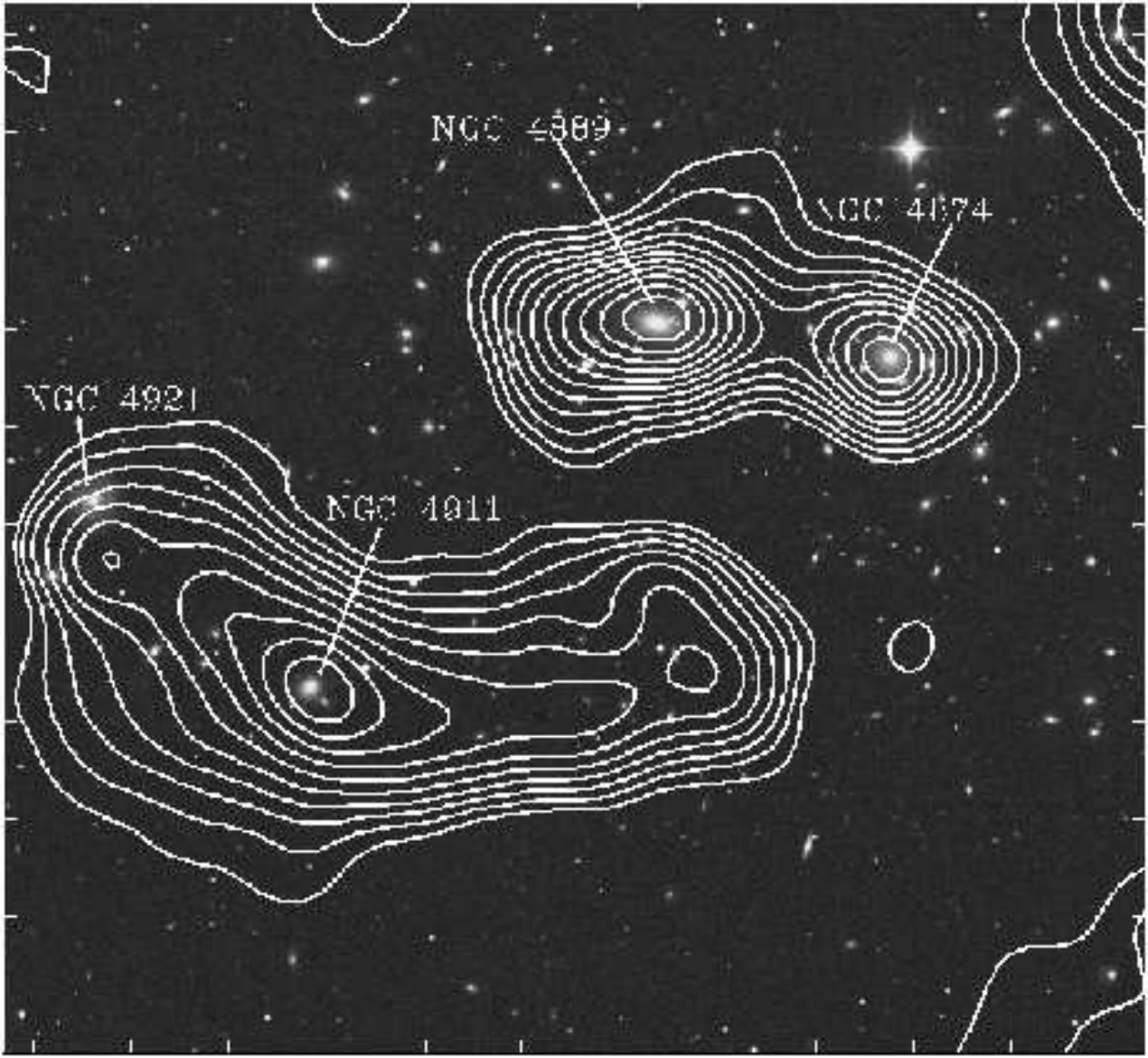,width=17cm}
\caption{The contours of the residuals superposed on the image of the DSS.
The contours are the same as in Fig.\ref{fig:resi} and Fig.\ref{fig:hr}.}
\label{fig:opt}
\end{figure*}

If we assume that the gas mass of the 
substructure  represents roughly 
ten percent of the total mass of the substructure, which is of the order of
magnitude of what is typically found (see for example David, Jones 
\& Forman \cite{david}), 
then we obtain for
the total mass
 M$_{tot,s} \sim 5-8 \times 10^{12}$ \msun. This is a few times larger than
 the mass of a massive galaxy and again somewhat at the low end of the mass of 
groups of
galaxies with typically a few $10^{13}$ \msun (e.g. Mulchaey \cite{mulchaey}).
NGC~4911 and NGC~4921 lie in this 
sub-structure and are by far the brightest galaxies (see Fig.\ref{fig:opt})
 in this
region. Dow \& White \cite{dow} give a value for the optical luminosity of 
NGC4911 of 
$log L_B $(in \lsun) $= 10.89$. Since NGC~4921 has a similar magnitude
 (see Doi et al. \cite{doi}), we can assume a similar 
luminosity. Assuming a mass-to-optical light ratio $\sim 10$, we obtain 
a total mass of NGC 4911 and NGC~4921 of $ \sim 10^{12}$ \msun, which is of
the same order of magnitude as the value required for the 
entire filamentary substructure. This indicates that the 
substructure is largely dominated by the two galaxies.
 
Matsushita (\cite{matsushita}) recently presented a correlation between mass
of the hot interstellar medium and optical luminosity $L_B$ for late type
galaxies. Applying this correlation to NGC~4921 and NGC~4911 we find a mass
for the hot phase in the order of $10^{10}$ \msun. This is smaller
than the value of a few times $10^{11}$ \msun. If correct, this implies that a 
large fraction of the gas comes indeed from an intra-group medium. 

The X-ray structure we see likely originates from gas in the surrounding 
intra-group medium of a small group of galaxies plus,
to some fraction, which we expect to be in ther order of a few up to 10 or
20 per cent, from the
stripped interstellar medium of the two galaxies, NGC~4911 and 
NGC~4921.

A cursory examination of Figure~\ref{fig:resi} would suggest that the principal direction of motion is to the West with perhaps an additional component to the North. Figure~\ref{fig:opt} shows the X-ray residual over the optical image from the DSS, indicating that NGC~4921 is at the Eastern end of the structure and NGC~4911 is right at the position where the structure bends. 
The only way we can explain this alignment is by assuming a direction of motion to the East, with NGC~4921 at the head of the structure and NGC~4911 behind. The different directions of the stripped tails may indicate different velocity directions for the galaxies, with NGC~4911 only having a projected velocity component towards the East while NGC~4921 has an additional projected component to
the North. This could indicate that the structure in which NGC~4911
and NGC~4921 are bound together changes its direction of velocity with time.

Are these two galaxies bound together? If they are, they should have similar measured velocities or redshifts. A combination of NASA Extragalactic Database (NED) and published data (Smith et al. \cite{smith}) indicates velocities of $cz=7973$~km$^{-1}$ and $cz=7560$~km$^{-1}$ for NGC~4911 and NGC~4921, respectively, suggesting that they are indeed gravitationally bound. (Note that the NED velocity for NGC~4921 is incorrect, having been superseded by the Smith et al. measurement.) Since the two galaxies have larger measured velocities than
the Coma cluster itself, it is reasonable to assume that the structure is 
located between us and the Coma cluster, thus lying somewhat in front of 
the Coma centre (although there is the small possibility that they are 
background).

We can estimate the temperature of the sub-structure  by 
assuming that it contributes about 30\% to the total flux in the region,
where it is located (see above). We perform simulations of spectra, assuming a 
main cluster temperature of 8~keV and a sub-group temperature of 1~keV and 
calculate the corresponding hardness ratio. We find a  
hardness ratio of 0.29, in agreement with the hardness ratio values in the region of the substructure.  If interactions 
occur and are important for the emission mechanism, we expect the cold phase 
to be at a temperature close to or below 1~keV. Thus, 1~keV seems to be a 
good upper limit for 
the temperature of the structure. Temperatures around or below 1~keV were 
found to be 
typical values for compact groups of galaxies (see for example Helsdon 
\& Ponman \cite{helsdon}). 

In the scenario described above, the two galaxies must have a tangential component with respect to the centre of the main cluster, which implies that the merger is not a simple two-body encounter.
The Coma cluster is surrounded by 
filamentary structures (see for example Huchra et al. \cite{huchra}), and their gravitational pull could explain the non-zero
impact parameter of the two galaxies on their infall onto the cluster.

\subsection{The two dominant central galaxies}

 Vikhlinin et al. (\cite{vikhlinin01}) have shown based on a study of {\it 
Chandra} data that the temperatures of the two dominant central galaxies NGC~4874 and 
NGC~4889 are significantly cooler than the surrounding gas. In our study we do 
not see a temperature drop in these regions, primarily because
we average over a too wide area in our analysis to resolve this 
cooler gas. 

\subsection{The X-ray deficit in the South and East of NGC~4911 and 
NGC~4921}

The X-ray deficit in the South and East is difficult to explain. Comparing
the overall morphology of the gas and temperature distribution with recent
studies based on hydrodynamical simulations, such as Roettiger et al. 
(\cite{roettiger}) and Ricker \& Sarazin (\cite{ricker}), we cannot find a 
scenario which matches all observed features. We therefore think that this 
decrement in X-ray emission is either due to tidal forces linked to the merger 
with the Western structure \footnote{in this case
the gas could follow the mass enhancement 
linked to the structure falling onto the centre from the West and provoking a 
diminuition of gas density in the East and South East}, or to an older merger, 
possible connected to the Northern part of the
Western substructure. This might be the remnant of a sub group, 
which already passed  the core once. However, the physics of this deficit
in emission might be more complex. We want to stress, again, that this 
decrement is 
not an artefact of our beta-model fitting. Changing the 
parameters for the model (position angle, core radius, centre position and 
background) over a considerable range does not alter the result
(see above).

\section{Conclusion}

We performed a substructure and spectroimaging study based on the \xmm data
of the Coma cluster. The data show the existence of substructures which
are very likely interacting with the main cluster. The combination of spatial and 
spectral information based on data obtained with \xmm  allows us to put  
global constraints on the dynamical state of the Coma cluster. These are
important for our understanding of the physics of clusters. The 
results presented here show the large potential of future studies of clusters
of galaxies observed with \xmm.

\begin{acknowledgements}

We would like to thank the \xmm EPIC-team for support. In particular we are
grateful to K. Dennerl, M. Freyberg and F. Haberl for fast and
very competent replies to our questions. Furthermore we would like to thank
J. Ballet for his help concerning the analysis based on the \xmm SAS software.
We are grateful for the discussions with S. Schindler concerning cluster 
mergers and M. Arnaud in general. We would also like to thank J. Huchra, who
acted as referee for the careful reading of the manuscript.

This research has made use of the NASA/IPAC Extragalactic Database (NED) which is operated by the Jet Propulsion Laboratory, California Institute of Technology, under contract with the National Aeronautics and Space  Administration.
 We acknowledge the use of NASA's SkyView facility 
(http://skyview.gsfc.nasa.gov) located at NASA Goddard Space Flight Center.
This research has made use of NASA's Astrophysics Data System Service.

This work is based on observations obtained with XMM-Newton, an ESA science
mission with instruments and contributions directly funded by ESA Member 
States and the USA (NASA).

\end{acknowledgements}

\end{document}